\documentclass[10pt,journal,compsoc]{IEEEtran}
\ifCLASSOPTIONcompsoc
  \usepackage[nocompress]{cite}
\else
  \usepackage{cite}
\fi
\ifCLASSINFOpdf
  \usepackage[pdftex]{graphicx}
\else
  \usepackage[dvips]{graphicx}
\fi

\usepackage{algorithmic}
\usepackage{array}

\ifCLASSOPTIONcompsoc
 \usepackage[caption=false,font=footnotesize,labelfont=sf,textfont=sf]{subfig}
\else
 \usepackage[caption=false,font=footnotesize]{subfig}
\fi
\usepackage{stfloats}

\ifCLASSOPTIONcaptionsoff
 \usepackage[nomarkers]{endfloat}
\let\MYoriglatexcaption\caption
\renewcommand{\caption}[2][\relax]{\MYoriglatexcaption[#2]{#2}}
\fi

\let\MYorigsubfloat\subfloat
\renewcommand{\subfloat}[2][\relax]{\MYorigsubfloat[]{#2}}
\usepackage{url}
\usepackage{subfiles}
\usepackage{amssymb}
\usepackage{amsmath,bm}
\usepackage{multirow}
\usepackage{adjustbox}
\usepackage{xcolor}
\usepackage{color}

\newcommand{\qyl}[1]{{\color{black}#1}}

\newcommand{\etal}{et al.}

\begin{document}

\title{LaplacianNet: Learning on 3D Meshes with Laplacian Encoding and Pooling}

%
%
%
%
\author{Yi-Ling Qiao,~
        Lin Gao,~
        Jie Yang,~
        Paul L. Rosin,~
        Yu-Kun Lai,~
        and~Xilin~Chen
        
\IEEEcompsocitemizethanks{
\IEEEcompsocthanksitem Y.-L Qiao, L. Gao and J. Yang are with the Beijing Key Laboratory of Mobile Computing and Pervasive Device, Institute of Computing Technology, Chinese Academy of Sciences, Beijing, China.
\IEEEcompsocthanksitem P.L Rosin and Y.-K Lai are with School of Computer Science and Informatics, Cardiff University, Wales, UK.
\IEEEcompsocthanksitem X. Chen is with Institute of Computing Technology, Chinese Academy of Sciences, Beijing, China.}
}
\IEEEtitleabstractindextext{%
\begin{abstract}
	3D models are commonly used in computer vision and graphics. With the wider availability of mesh data, an efficient and intrinsic deep learning approach to processing 3D meshes is in great need. Unlike images, 3D meshes have irregular connectivity, requiring careful design to capture relations in the data. To utilize the topology information while staying robust under different triangulation, we propose to encode mesh connectivity using Laplacian spectral analysis, along with Mesh Pooling Blocks (MPBs) that can split the surface domain into local pooling patches and aggregate global information among them. We build a mesh hierarchy from fine to coarse using Laplacian spectral clustering, which is flexible under isometric transformation. Inside the MPBs there are pooling layers to collect local information and multi-layer perceptrons to compute vertex features with increasing complexity. To obtain the relationships among different clusters, we introduce a Correlation Net to compute a correlation matrix, which can aggregate the features globally by matrix multiplication with cluster features. Our network architecture is flexible enough to be used on meshes with different numbers of vertices. We conduct several experiments including shape segmentation and classification, and our LaplacianNet outperforms state-of-the-art algorithms for these tasks on ShapeNet and COSEG datasets. 
\end{abstract}

}

\maketitle

\IEEEdisplaynontitleabstractindextext

%
\IEEEpeerreviewmaketitle

\IEEEraisesectionheading{\section{Introduction}\label{sec:introduction}}

%
%
%
%
\IEEEPARstart{A}{nalyzing}
 high-quality 3D models is of great significance in computer vision and graphics. \qyl{Better understanding of 3D shapes would benefit many tasks such as segmentation, classification and shape analysis.} Recently, deep learning methods have been prevalent in 2D image processing tasks such as image classification \cite{karpathy2014large,chen2017reference} and semantic segmentation~\cite{long2015fully,kwak2017weakly}. With the help of large-scale image datasets and improved computational resources, deep learning methods boost performance of image processing algorithms by a large margin. \qyl{Inspired by the success in images, researchers also apply learning algorithms to 3D data.}

Recently, large-scale 3D datasets have made it possible to train neural networks for 3D shapes. Nevertheless, it is not a simple extension to apply neural networks in the 3D space. There are various 3D representations. The majority of 3D representations such as meshes, point clouds etc. are non-canonical, requiring special design to put them through neural networks. To address this, some approaches are trained on ModelNet~\cite{wu20153d} and deal with voxels, but the resolution of voxel data is limited due to the curse of dimensionality. Alternatively, point clouds representing an object by a set of unstructured points with their $xyz$ coordinates are commonly used. 
\qyl{However, point clouds do not carry connectivity information and therefore are less efficient than meshes to represent shapes, and may have ambiguities when two surfaces are close.}
Another representation, the 3D mesh, is a fundamental data structure in computer graphics and vision, which not only encodes geometry but also topology
and therefore has better descriptive power than the point cloud.
A mesh is a graph with vertices, edges and faces that characterize the surface of a shape. For deep learning methods, mesh data is more compact but irregular when compared to voxels, making simple operations in the image domain such as convolutional kernels highly non-trivial. It also contains richer structure than a point cloud, which can be exploited when learning on 3D meshes.
\qyl{This paper aims to propose a flexible network structure that can perceive connectivity information while staying robust under different triangulation.} 

\begin{figure*}[t]
	\begin{center}
		\includegraphics[width=\linewidth]{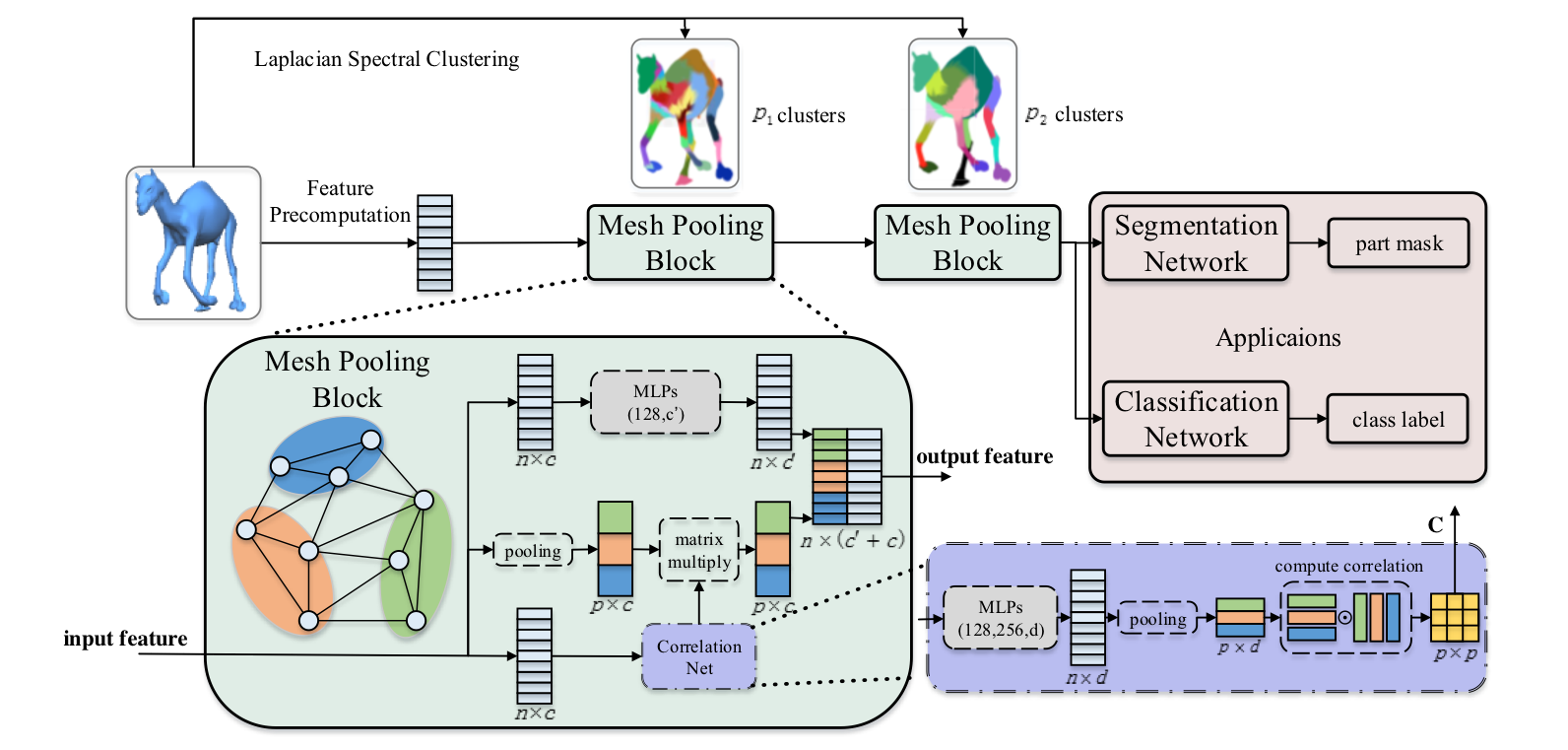}
	\end{center}
	\caption{Overview of our Laplacian Pooling Network (LaplacianNet). Given a 3D mesh as input with the aim of producing some kind of vertex function (depending on the application) as output, the pipeline of our network has three main components. First we preprocess the mesh to compute Laplacian eigenvectors and spectral clustering. These along with vertex coordinates and normals form the input feature to the network. 
	Second, we stack several Mesh Pooling Blocks (MPBs) to analyze the shape model under multiple resolutions. \qyl{An MPB includes some multi-layer perceptrons (MLP) to compute features independently for each vertex, and also a pooling layer to compute the features for clusters. A Correlation Net learns a correlation matrix $\bm{C}$ to fuse features across clusters. Then the fused features are concatenated with MLP results, 
	and then associated with individual vertices in the cluster.}
	After a sequence of MPBs (two in this illustration), 
	    the features are fed into Application Network to produce output according to specific applications.}
	\label{fig:networkstruct}
\end{figure*}

\qyl{
To learn on 3D meshes, we propose the Laplacian Encoding and Pooling Network (LaplacianNet), which takes raw features of mesh models as input and outputs a function defined on the vertices. Inspired by image processing networks, we observe an intuitive principle that vertex features should be computed independently and associatively in different parts of the network. We therefore extend this ideology into non-Euclidean space of 2-manifolds. In our design as in Figure~\ref{fig:networkstruct}, the basic network structure involves consecutive Mesh Pooling Blocks (MPBs). Each block can split the surface into patches, like super-pixels in the image domain, by Laplacian spectral clustering. After splitting, the MPB can simultaneously compute features of individual vertices and clusters. Considering the relationships between clusters, we use a Correlation Net to compute a matrix that can fuse the information globally. Compared to images, a major difficulty for learning on meshes is that the vertices are unordered, and so are the clusters. For this reason, a fully-connected layer cannot work out the correlation matrix effectively. Therefore, we disentangle the correlation by independently mapping the clusters into a vector space. Then, the correlation between a pair of clusters is determined by the inner product of their corresponding vectors.
}


Before our LaplacianNet, effort has been made to learn on 3D meshes. Please refer to~\cite{bronstein2017geometric} for history and frontier research of deep learning on geometric data. Some works exploit geodesic distances on shapes~\cite{masci2015geodesic,boscaini2016learning}. Based on this metric they design a spatial convolutional operator. \qyl{Geodesic distance could be invariant under isometric transformation, making their networks more robust. Compared to their directly computing the distance, our network utilizes the geodesic distance in an implicit way. The clustering is performed in the Laplacian feature domain, where Euclidean distance can approximate geodesic distance on the manifold~\cite{Crane:2017:HMD}. As a result, the pooling could stay robust under isometric deformation and different triangulation (shown in Figure~\ref{fig:remesh}).} Others choose to work in the spectral domain~\cite{2013spectral,li2018adaptive}, by defining convolutional kernels in the Fourier domain. However, the dependency of coefficients on the domain basis makes it difficult to share weights across shapes. Consequently, works like~\cite{yi2017syncspeccnn} have modules purposefully designed to synchronize the basis of domains. \qyl{Our network does not suffer from changing domains, and we also propose a flexible structure, Correlation Net, to address the alignment of clusters across models. Compared to all the aforementioned methods, we use both spectral and spatial information, such that our network can utilize the connectivity of meshes while staying robust under different domains with inconsistent triangulation. } 

The pipeline of our approach is shown in Figure~\ref{fig:networkstruct}, which learns cross-domain mesh functions using both spatial and spectral information. Overall, LaplacianNet has the following three modules: 1) the preprocessing step computes vertex features and clusters from the raw mesh data; 2) \qyl{the Mesh Pooling Block (MPB) calculates local features within local regional clusters and collects global information through Correlation Net;} 3) the last part of the network depends on the specific application, e.g., it may output segmentation masks or classification labels.

In the experiments, we first evaluate the importance of mesh pooling blocks and the choice of the input features. We then justify that our single network can deal well with different mesh models with different numbers of vertices. To test the capability of our network, we train our network on the ShapeNet and COSEG datasets to perform classification and segmentation tasks, which are fundamental shape understanding tasks in computer vision, and show superior overall performance.

The main contributions of our method are as follows:

\begin{itemize}
	\item \qyl{We propose the Laplacian Encoding and Pooling Network (LaplacianNet), a general network for learning on 3D meshes, which can utilize the connectivity of meshes while staying robust under different triangulation. }
	\item \qyl{We propose a flexible pooling strategy that can split model surfaces into clusters, like superpixels in images. By varying the clusters from fine to coarse, the network can process meshes hierarchically.}
	\item \qyl{We introduce a Correlation Net to compute the relationship among clusters. The computation process circumvents the randomness of cluster ordering, enabling consistency across domains.}
\end{itemize}

\section{Related Work}\label{sec:relatedwork}

We first summarize deep learning methods on 3D representations, and then provide a brief introduction to alternative input shape features. Finally, we review recent methods for mesh segmentation, which is a fundamental task when analyzing shapes. 

\noindent\textbf{3D Deep Learning.} With the increasing availability of 3D models, deep learning methods for analyzing 3D data structure have been widely studied nowadays. There are several representations for 3D shapes, including voxels, point clouds and meshes. The voxel representation is similar to pixels in the 3D space, which can utilize a direct extension of 2D convolutional networks~\cite{qi2016volumetric}. The point cloud is intensively researched, with works like \cite{qi2017pointnet} learning the transformation matrix of points and obtaining decent results on multiple datasets. {Qi et al.~\cite{qi2017pointnet++} further propose PointNet++ that adds pooling operations, where pooling areas are selected by nearest neighbors. 
	We aim at different problems: While they address problems on point clouds, our method focuses on meshes. To better understand 3D meshes, the topology information is deeply exploited in our method. First, we use topology to perform clustering. In contrast, the nearest neighbor pooling strategy of PointNet++ cannot perceive surface structure of meshes. Second, we use features that contain topology information as input. Meanwhile, we believe that it is not appropriate to carry the entire topology during the whole process. The connectivity information is irregular, large, hard to compute, and sensitive to noise. We instead condense the topology information by encoding the connectivity into the pooling areas and input features.}

For the mesh representation, spatial methods~\cite{masci2015geodesic,boscaini2016learning} define convolutional kernels on surfaces. 
\qyl{Our LaplacianNet also utilizes spatial information, as the pooling strategy partitions the surface into patches, acting like super-pixels in image processing.}
For spectral methods, Bruna~\etal~\cite{2013spectral} introduce a spectral convolutional layer on graphs, which can be interpreted as convolutions in the Fourier domain. 
Henaff~\etal~\cite{henaff2015deep} handle large-scale classification problems. They propose to exploit the underlying relationships among individual data points by constructing a graph of the dataset and then solving a graph-based classification. When processing the graph, they introduce a CNN structure with spectral pooling. Compared with ours, their work is interested in each node and the pooling cluster does not contain geometric information.
A fundamental problem of spectral convolution is  generalizing across domains since the coefficients of spectral filters are basis dependent~\cite{bronstein2017geometric}. To address this problem, Yi~\etal~\cite{yi2017syncspeccnn} further propose SyncSpecCNN to perform convolutional operations in a synchronized spectral domain, where they train a Spectral Transformer Network to align the functions in different domains. Compared to all the spectral CNN methods above, our LaplacianNet takes advantage of Laplacian spectral analysis to encode mesh topology and identify a spatial pooling strategy but avoids suffering from its dependency on the domain. {We also show in the Experiments section that our method is robust under different object categories, number of vertices, and triangulation.}

\noindent\textbf{Shape Features.}
Over the years, many shape features have been developed to describe shape characteristics, including curvatures, geodesic shape contexts, geodesic distance features as used in \cite{kalogerakis2010learning}, Heat Kernel Signatures~\cite{sun2009concise}, Wave Kernel Signatures~\cite{aubry2011wave}, etc.
Our LaplacianNet exploits mesh Laplacian spectral analysis, which provides an effective encoding of mesh topology. Laplacian eigenvectors also help us to cluster vertices for pooling layers, and it is an intrinsic feature describing the geometry of shapes. Readers can refer to \cite{von2007tutorial} for details about computation and applications of graph Laplacian. However, two essential problems with Laplacian eigenvectors are that the sign is not well-defined, and perturbation occasionally occurs in high frequency terms. To eliminate these ambiguities, we use the absolute value of low frequency terms as input. 


\noindent\textbf{Mesh Segmentation.}
Mesh segmentation has long been a fundamental task in the field of computer vision and graphics. There are unsupervised and supervised methods to perform this task. For unsupervised methods, recent work usually uses the correspondence or correlation between shape units to co-segment a collection of objects in the same category~\cite{wu2013unsupervised,shu2016unsupervised,huang2011joint,sidi2011unsupervised}. Those methods essentially analyze a whole dataset of similar 3D shapes and cluster shape parts that can be consistently segmented into one class. 
Other works try to take advantage of labeled data to develop a supervised method. Thanks to recent shape segmentation datasets~\cite{chen2009benchmark,yi2016scalable,wang2012active,shilane2004princeton}, supervised methods obtain higher accuracy than unsupervised ones. Among all those datasets, COSEG~\cite{yi2016scalable} and ShapeNet~\cite{yi2016scalable} have sufficiently many samples to train a network with reasonable generalizability, so we conduct experiments on the two datasets. Previous deep learning  methods usually design different architectures to do segmentation. For example, George et al.~\cite{george20183d} design a multi-branch 1D convolutional network and Wang et al.~\cite{wang20183d} put convolutional kernels on neighboring points and a pooling layer on the coarsened mesh. A 2D CNN is embedded into a 3D surface network by a projection layer in \cite{kalogerakis20173d}. 
Guo et al.~\cite{guo20153d} concatenate different feature vectors into a rectangular block and apply CNNs to this image-like domain. 
Our method aims to develop a general network for learning on 3D meshes, and we demonstrate that our general approach outperforms existing methods in most cases.

\section{Methodology}\label{sec:methodology}

\subsection{Problem Statement}
Our proposed network is a general method for 3D meshes, 
\qyl{capable of dealing with different numbers of vertices.} Suppose that the current input  $\mathcal{G}=(\mathcal{V},\mathcal{E})$ is a mesh with $N$ vertices. 
There is an input feature function $f$ defined on vertices $\mathcal{V}$, i.e., $f:\mathcal{V}\rightarrow\mathbb{R}^{N\times c}$ where $c$ is the dimension of input features, typically containing coordinates, normals, mesh Laplacian, curvatures, etc. 
At the same time, there is a target function $g$ we aim to produce. It can usually be a vector function such as a segmentation $g:\mathcal{V}\rightarrow {C_s}^N$ where each entry corresponds to the label of a vertex,  or a single value like classification category for the whole object $g:\mathcal{G}\rightarrow C_l$, where $C_s$ and $C_l$ are the sets of segmentation labels and classification categories respectively. We would like to mention that $g$ may also represent other functions including texture or normal.
Our aim is to design a general neural network that learns the mapping from input feature $f$ to the output $g$. For the segmentation and classification tasks, we precompute the normal and mesh Laplacian eigenvectors as input features. Our network will output an $N\times |C_s|$ matrix for segmentation, which gives the score for each vertex belonging to each segmentation label, or a $|C_l|$ dimensional softmax vector for classification. 

\subsection{Feature Precomputation}\label{sec:feapre}


We aim to use minimal features to characterize local geometric information. For this purpose, we precompute the vertex normals and mesh Laplacian eigenvectors, which along with vertex positions are used as the input to our network. The mesh Laplacian provides an effective way to encode mesh connectivity. 

For later pooling layers, we use Laplacian spectral clustering~\cite{von2007tutorial} at multiple resolutions. Different from spectral convolutions in~\cite{henaff2015deep}, our pooling layers reduce \qyl{any number of vertices to a desired dimension}, which makes it possible for our method to cope with meshes with different topology. 
In practice, the Laplacian matrix is computed by
\begin{equation}
	\bm{L}=\bm{A}^{-1}(\bm{D}-\bm{W})
\end{equation}
where $\bm{A}=diag(a_1,...,a_n)$ are vertex weights defined as local Voronoi areas $a_i$, equal to one third of the sum of one-ring triangles areas. $\bm{W}=\{w_{ij}\}_{i,j=1,...,N}$
is the sparse cotangent weight matrix, discretization of continuous Laplacian over mesh surfaces~\cite{Meyer2003}, and $\bm{D}$ is the degree matrix which is a diagonal matrix with diagonal entries $d_{ii}=\sum_{j=1}^{N}{w_{ij}}$. The Laplacian feature $\bm{\Phi}$ is the eigenvectors of matrix $\bm{L}$. Please refer to~\cite{von2007tutorial} for detailed computation and applications of graph Laplacian. 

To cluster vertices at different levels, we perform k-means clustering
on $\bm{\Phi}$ with different numbers of clusters  $k=p_l$ such that vertices are clustered into $p_l$ clusters for the $l^{\rm th}$ pooling block. $l=1, 2, \dots, L$, and $L$ is the number of levels. \qyl{To achieve local-global feature extraction, $p_l$ decreases as $l$ increases.} \qyl{Note that since the clustering is in the feature domain, vertices on the surface within one cluster are not necessarily connected. This is also reasonable because some semantically similar vertices can be far away.}

\subsection{Network Architecture}
The architecture of our LaplacianNet is illustrated in Figure~\ref{fig:networkstruct}.

Given the preprocessed feature function $f$ defined on vertices $\mathcal{V}$, our LaplacianNet takes the feature matrix as input, each row being a feature vector of a vertex. 
By reducing the input features to a matrix, we avoid the complex graph structure and make it tractable for neural networks.

\begin{figure*}[t]
	\begin{center}
		\includegraphics[width=\linewidth]{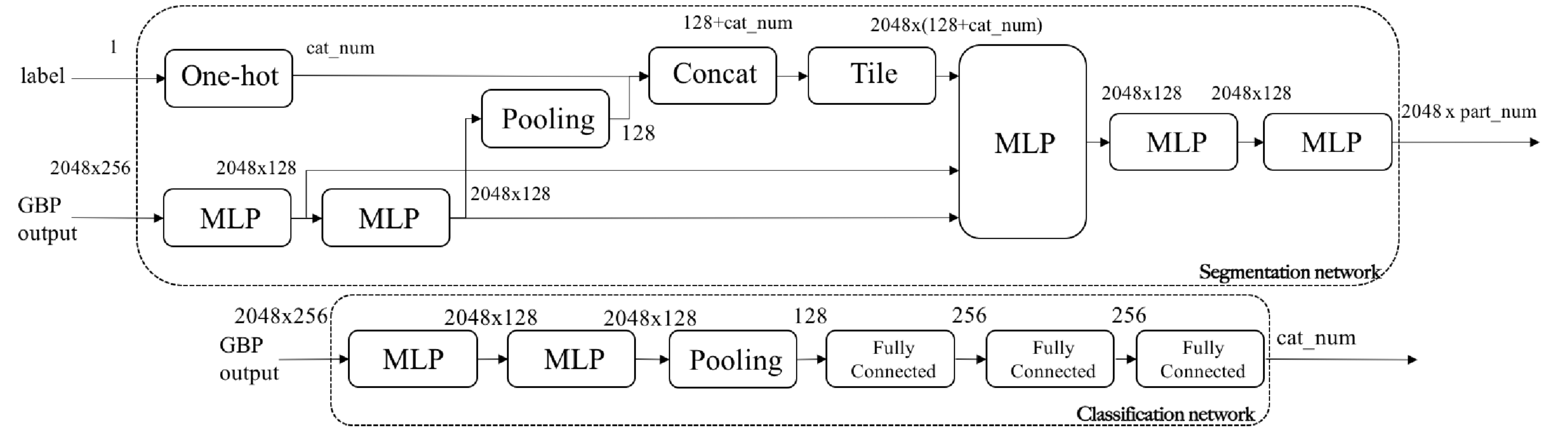}
	\end{center}
	\caption{The application network for segmentation and classification. For the segmentation task (top row), the application specific network takes as input the category label of a certain item and the features defined on vertices. The vertex features go through two MLP layers and then are duplicated into two branches, one of which is sent to a global pooling, combined with the one-hot vector of input label and then attached back to vertex features from the other branch. Finally, two MLP layers are used to compute the final segmentation mask. For the classification network (bottom row), the vertex features sequentially go through MLP layers, global pooling, and fully connected layers. Eventually a softmax vector for candidate labels is predicted.}
	\label{fig:appnet}
\end{figure*}

Several mesh pooling blocks (MPBs) are then applied in various resolutions for multiple times. At the end of MPB blocks, the application network outputs the target function $g$. Details about pooling blocks will be further discussed in the next section. The architecture of the application network for classification and segmentation is shown in Figure~\ref{fig:appnet}. 

In our design, to circumvent the complex and irregular topology of mesh data, we seek a pipeline that can concisely describe the relationship among vertices. Instead of directly processing edges
$\mathcal{E}$, we simplify this problem by only processing vertices $\mathcal{V}$ of mesh $\mathcal{G}$. Nevertheless, edges are not ignored, but instead implicitly encoded into Laplacian eigenvectors and spectral clusters as described in the previous subsection. Since the mesh Laplacian is intrinsically induced from geodesic distances, our method is robust to remeshing and isometric transformations.

Moreover, since the total number of vertices in a mesh can vary significantly from model to model, an ideal network architecture should be able to deal with meshes with different numbers of vertices.
Our solution is to design mesh pooling blocks, which turn meshes of arbitrary sizes into levels with the same number of clusters. By stacking together several blocks in a multi-resolution manner, our network can learn to extract useful features from the mesh.
In addition, our network uses parameters more effectively with shared weight Multi-Layer Perceptron (MLP), which also helps avoid over-fitting by reducing the complexity of our network. Our method consequently achieves good results for shape classification and segmentation.



\subsection{Mesh Pooling Blocks}\label{sec:pooling}
\qyl{A mesh pooling block is composed of three modules: the Multi-Layer Perceptron (MLP) layers~\cite{rumelhart1985learning}, the pooling layers, and a Correlation Net. Figure~\ref{fig:networkstruct} shows an illustration of a mesh pooling block. Each mesh pooling block obtains its input feature $\bm{F}_l$ of size $N\times c_l$ from the previous layer. It also gets the cluster mask $\bm{M}_l$ from the precomputation step, where an entry $m_{l,i} \in \{1, 2, \dots, p_l\}$ in the mask $\bm{M}_l \in \mathbb{Z}^N$ indicates for node $i$ which cluster it belongs to. The total number of clusters for the $l^{\rm th}$ block is $p_l$. }

\qyl{As illustrated in Figure~\ref{fig:networkstruct}, the data flows through three branches. The upper path is a series of MLP layers, learning vertex features of increasing complexity; the middle path is the pooling layer followed by a correlation matrix multiplication, which fuses global and local information; the bottom path is a Correlation Net that computes the correlation matrix, learning the interaction among clusters.}

\qyl{The upper path of the MPB in Figure~\ref{fig:networkstruct} is a set of MLP layers with shared weight perceptrons connected to all the vertices. For a certain vertex $i$, an MLP layer multiplies its input $\bm{F}_{l,i}$ and weight matrix $\bm{W}$ with bias $\bm{b}$, followed by a $ReLU(\cdot)$  activation function. The operation of this layer can be expressed as }
\begin{equation}
	MLP(\bm{F}_{l,i})=ReLU(\bm{WF}_{l,i}+\bm{b})
\end{equation}

For the pooling layer, the input includes features $\bm{F}_l$ from the last block and a cluster mask $\bm{M}_l$. Its result is defined as applying the operation to all the nodes belonging to the same cluster.  
Take max pooling as an example. The pooling result $\bm{P}_{l,j}$ corresponding to the $j^{\rm th}$ cluster of the $l^{\rm th}$ block is:
\begin{equation}
	\bm{P}_{l,j}=\max_{m_{l,i}=j}\bm{F}_i
\end{equation}

\qyl{Furthermore, we want the features to be computed across clusters. 
For images, convolutional kernels can be used to fuse pooling results. 
However, since triangle meshes do not have a regular grid and consistent clustering, 
such an approach does not work.
Standard global pooling can be a simple choice, but each cluster has equal contribution and detailed information is lost. 
At the bottom path in Figure~\ref{fig:networkstruct}, to aggregate information from all clusters, we multiply the pooling results with a correlation matrix $\bm{C}_l=\{c_{ij}\}_{p_l\times p_l}$. Each entry $c_{ij}$ measures the correlation between the $i^{\rm th}$ and $j^{\rm th}$ clusters, such that the aggregated pooling result is obtained as $\tilde{\bm{P_l}}=\bm{C_l}\bm{P_l}$. The Correlation Net computes the matrix $\bm{C}$ by learning latent vector embedding for each cluster:}
\begin{equation}
\bm{\Psi}_l=\{\psi_{lj}\}=Pooling(MLP(\bm{F}_{l})),
\end{equation} 
 and entries of the correlation matrix are inner products of latent vectors of cluster pairs $c_{lij}=<\bm{\psi}_{li},\bm{\psi}_{lj}>$.
\begin{equation}
	\tilde{\bm{P}}_l = \bm{\Psi}_l\bm{\Psi}_l^{\bm{T}}\bm{P}_l
\end{equation}


The concatenation layer combines vertex features from the upper-path MLPs and aggregated pooling results. For the vertex $i$ in cluster $j$, its output feature is written as
\begin{equation}
	\{MLP(\bm{F}_{l,i}), \tilde{\bm{P}}_{l,j}\}
\end{equation}

\qyl{In summary, the MPB is used for both processing the local information and finding the relationships among local patches.
The motivation for pooling in the mesh is to better understand spatial information. Ideally, the input vertices are hierarchically clustered into different areas, and the relationships between areas need to be considered. In practice, we process the hierarchical information by clustering vertices into different numbers of clusters. The Correlation Net learns to compute a correlation matrix to describe the relative relationships among spatial areas after pooling. 
\qyl{Using Laplacian clustering ensures that the clustering is robust under different triangulation. Also, the Euclidean distance in the Laplacian feature domain approximates the geodesic distance~\cite{Crane:2017:HMD}. Therefore, such a clustering can find meaningful patches in the spatial domain.}
Moreover, the clustered areas give a reasonable partition, as shown in the camel example in Figure~\ref{fig:networkstruct} where vertices with the same color roughly belong to one semantic area.
}

\section{Experiments}\label{sec:experiments}

\subsection{Implementing Details}
We implement our method using Tensorflow. We train and test the network on a desktop with an NVIDIA GTX1080 GPU. The optimization is achieved using the Adam~\cite{kingma2014adam} solver with learning rate $7\times 10^{-4}$ and momentum $0.9$. The network is trained for 200 epochs with batch size 8. The input features have 22 dimensions, including 6 dimensions of positions and normals, and the other 16 dimensions are the absolute values of the Laplacian eigenvectors corresponding to the 16 lowest frequencies. According to the experimental results in the next section, by default we choose to use two Mesh Pooling Blocks with 16 and 8 clusters respectively. There are two MLP layers in a mesh pooling block outputting 128 and 256 channels. \qyl{Following MPBs is a specific application network based on the task to be performed. In total, the network's depth is $11$ for the segmentation and classification tasks.}


\subsection{Network Evaluation}
We now evaluate different parts of our network. This series of evaluation is conducted on the COSEG dataset, with results  shown in Tab.~\ref{tab:coseg}.

First we test the usefulness of the input features. 
The performance of a certain algorithm can be affected by the features used. In the experiments, we use coordinates, normals and Laplacian eigenvectors as vertex features. In this part, we test the network without one of those three features. We can see that a combination of all the features achieves the best results.

Second, we test the setting of Mesh Pooling Blocks. By default, our method uses 2 MPBs. We compare this with alternative numbers of MPBs ranging from 0 to 3. The results show that 2 MPBs (our default method) gives the best performance. \qyl{We also test the usefulness of our Correlation Net. Ours-noCorrNet is the network without the Correlation Net and matrix $\bm{C}$. We observe that the the aggregation of global information is important for our network. }

\begin{figure}[t]
	\begin{center}
		\includegraphics[width=\linewidth]{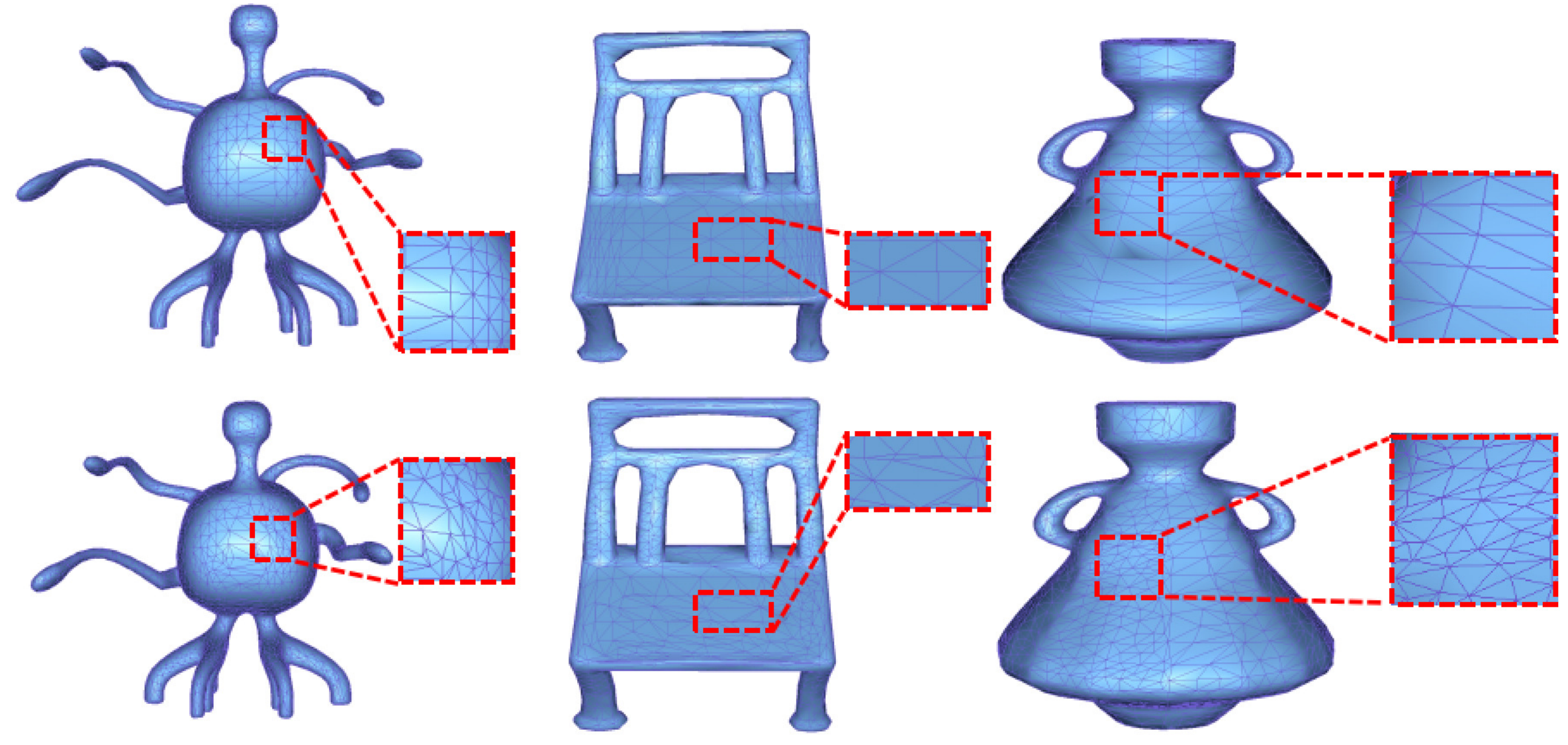}
	\end{center}
	\caption{Robustness to changing connectivity. In this experiments, we change the connectivity of meshes from the COSEG datasets, and test whether our network can consistently perform well. The first row is the original objects, while the meshes in the second row are remeshed into more irregular triangulation. As shown in Table~\ref{tab:coseg}, the segmentation accuracy still remains high. }
	\label{fig:remesh}
\end{figure}

Third, another difficulty when handling mesh data is varying mesh connectivity. As mentioned before, our method is robust under different mesh triangulation as a result of the pooling strategy. The Laplacian eigenfunction is induced from geodesic distances and therefore invariant under isometric transformation, so the pooling areas as well as input features can essentially stay unchanged when we remesh the object. We visualize the connectivity of the objects before and after remeshing in Figure~\ref{fig:remesh}, which is obtained by subdivision followed by mesh decimation~\cite{Garland1997} to generate irregular connectivity. Moreover, the quantitative results (Ours-remesh) in Tab.~\ref{tab:coseg} show that our network is robust under different triangulation.

Fourth, our network does not rely on the same vertex numbers for models.  An experiment is performed to test this. In this experiment, we first simplify COSEG models to 1500, 2000 and 2500 vertices. Then we split the training and test set in the same strategy for all three resolutions. We train our network on a mix of two of the datasets 
and test models from the third (Ours-1500, Ours-2000 and Ours-2500 in the table). Accuracies in Tab.~\ref{tab:coseg} show that our network works well with varying vertex numbers, compared to the last row where LaplacianNet is trained with models all containing 2048 vertices.

\subsection{Part Segmentation}\label{sec:seg}
In this section, we use MPBs to conduct part segmentation on the ShapeNet~\cite{yi2016scalable,chang2015shapenet} and COSEG~\cite{wang2012active} datasets. The application network for segmentation is illustrated in Figure~\ref{fig:appnet}(top row). Its input is the category label and features from the previous MPB. The output is a softmax score for each category. The application network has two perceptron layers, a maxpooling layer and two fully connected layers. We minimize the cross-entropy loss between the one-hot vector of ground truth and the network output. 

ShapeNet is a large repository of shapes from multiple categories. To leverage this dataset to perform segmentation, Yi~\etal~\cite{yi2016scalable} develop an active learning method for efficient annotation. However, their annotations are not directly on the mesh vertices, but on the point cloud resampled from the shapes. To recover the graph structure of manifold surfaces for computation of mesh Laplacian and segmentation, we apply \cite{huang2018robust} on the original ShapeNet models. After that, we transfer the annotations on the point clouds to 
the nearest mesh vertices.

\begin{figure}[t]
	\begin{center}
		\includegraphics[width=\linewidth]{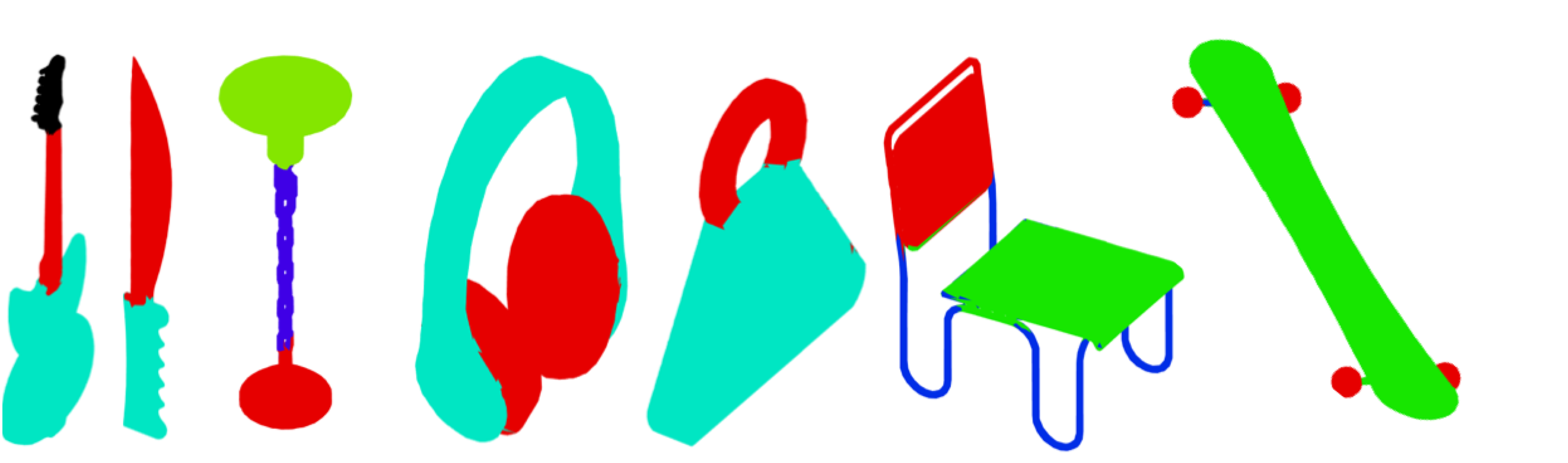}
	\end{center}
	\caption{Qualitative results on ShapeNet. The segmentation results produced by our method are plausible. }
	\label{fig:qualitative}
\end{figure}

Two metrics are used in previous segmentation results on ShapeNet, namely accuracy and IoU (Intersection-over-union). We compute both of them to compare with state-of-the-art deep learning methods on 3D shapes~\cite{yi2017syncspeccnn,kalogerakis20173d}, and some other methods performing segmentation on ShapeNet based on point clouds such as \cite{verma2018feastnet}. Tab.~\ref{tab:shapenet} shows that our method achieves the highest average accuracy, and outperforms state-of-the-art methods on 13 out of 16 categories. In terms of IoU, our performance is comparable to the state-of-the-art, achieving the best performance in 8 categories. Some segmentation results are presented in Figure~\ref{fig:qualitative}.

\begin{table}[h]
	\begin{center}
		\caption{Segmentation accuracy on COSEG. We compare with \cite{xie20143d} and \cite{wang20183d} in the first two rows. In the last row, LaplacianNet is trained on models with 2048 vertices. To test the robustness on different vertex, \qyl{we simplify COSEG models to 1500, 2000 and 2500 vertices respectively. 
		We train three networks on two of the three datasets but test on the third. Ours-1500, Ours-2000 and Ours-2500 show the accuracy when the test set has 1500, 2000 and 2500 vertices.}
		 We observe that our network performs similarly well with different vertex numbers. 
			Stable performance is also obtained when applying our method to remeshed models with more irregular connectivity (Ours-remesh). The ablation test on the features shows that all three kinds of features  contribute to the performance. 
			We also vary the number of MPBs
			\qyl{and find that using two MPBs performs best}. In general, our method achieves state-of-the-art results in all three categories.}
		\label{tab:coseg}
		\begin{tabular}{c|ccc}
			\hline
			&Chair&Vase&Tele-alien \\ \hline
			Xie~\etal~\cite{xie20143d}&87.1\%&85.9\%&83.2\% \\ 
			Wang~\etal~\cite{wang20183d}&\textbf{95.9}\%&91.2\%&93.0\%\\ \hline \hline
			Ours-noMPB&76.6\%&77.8\%&80.7\% \\ 
			Ours-1MPB&85.3\%&86.1\%&88.9\% \\ 
			Ours-3MPBs&90.1\%&\textbf{92.2}\%&91.6\% \\ 
			Ours-noCorrNet&86.9\%&85.6\%&90.7\% \\  \hline
			Ours-1500&90.3\% &90.0\% & 89.0\% \\ 
			Ours-2000&90.9\% &91.6\%  &89.3\%  \\ 
			Ours-2500&87.0\%  &86.6\% &88.5\%  \\ 
			Ours-remesh& 92.1\% & 91.5\% & 91.8\% \\ \hline
			Ours-noCoordinates&90.6\%&88.6\%&84.2\% \\ 
			Ours-noNormal&87.6\%&86.1\%&85.0\% \\  
			Ours-noLaplacian&79.6\%&87.1\%&86.1\% \\ \hline
			Ours&94.2\%&\textbf{92.2}\%&\textbf{93.9}\% \\  \hline
		\end{tabular}
	
	\end{center}
\end{table}

\begin{table*}[t]
	\begin{center}
			\caption{Accuracy and IoU of different methods on ShapeNet. For the task of 3D shape segmentation, we compare our method with Shapeboost~\cite{kalogerakis2010learning}, Guo~\cite{guo20153d}, and ShapePFCN~\cite{kalogerakis20173d} in the metric of accuracy. And we compare with FeaStNet~\cite{verma2018feastnet}, ACNN~\cite{boscaini2016learning}, Yi~\cite{yi2017syncspeccnn}, and  VoxelCNN~\cite{yi2017syncspeccnn} in the IoU (Intersection-over-union) metric. Our LaplacianNet achieves highest accuracy.
		}	
		\label{tab:shapenet}
	\begin{adjustbox}{max width=\textwidth}
		\begin{tabular}{c|p{0.6cm}|p{0.5cm}p{0.5cm}p{0.5cm}p{0.5cm}p{0.5cm}p{0.5cm}p{0.5cm}p{0.5cm}p{0.5cm}p{0.5cm}p{0.5cm}p{0.5cm}p{0.5cm}p{0.5cm}p{0.5cm}p{0.5cm}}
			\hline
			\multirow{2}{*}{Method}
			&mean&plane&bag&cap&car&chair&ear&guitar&knife&lamp&laptop&motor&mug&pistol&rocket&skate&table \\
			&&&&&&&phone&&&&&bike&&&&board& \\ \hline
			Shapeboost (acc)&77.2&85.8&93.1&85.9&79.5&70.1&81.4&89.0&81.2&71.1&86.1&77.2&94.9&88.2 &79.2&91.0&74.5\\ Guo (acc)&77.6&87.4&91.0&85.7&80.1&66.8&79.8&89.9&77.1&71.6&82.7&80.1&95.1&84.1&76.9&89.6&77.8 \\
			ShapePFCN (acc)&85.7&\textbf{90.3}&\textbf{94.6}&\textbf{94.5}&90.2&82.9&\textbf{84.9}&91.8&82.8&78.0&95.3&\textbf{87.0}&96.0&91.5&81.6&91.9 &84.8\\ \hline 
			ours (acc) &\textbf{91.5}&89.6&90.2&88.2&88.2&\textbf{83.2}&82.3&\textbf{95.6}&\textbf{88.7}  &\textbf{87.4}&\textbf{96.3}&70.6&\textbf{97.0}&\textbf{92.7}&\textbf{82.2}&\textbf{94.7}&\textbf{92.6} \\ 
			\hline\hline
			FeaStNet (IoU)&81.5&79.3&74.2&69.9&71.7&87.5&64.2&90.0&80.1&78.7&94.7&62.4&91.8&78.3&48.1&71.6&79.6 \\
			ACNN (IoU)&79.6&76.4&72.9&70.8&72.7&86.1&71.1&87.8&82.0&77.4&95.5&45.7&89.5&77.4&49.2&82.1&76.7 \\
			VoxelCNN (IoU)&79.4&75.1&72.8&73.3&70.0&87.2&63.5&88.4&79.6&74.4&93.5&58.7&91.8&76.4&51.2&65.3&77.1 \\
			Yi (IoU)&\textbf{84.7}&81.6&81.7&\textbf{81.9}&75.2&\textbf{90.2}&\textbf{74.9}&\textbf{93.0}&\textbf{86.1}&\textbf{84.7}&\textbf{95.6}&\textbf{66.7}&92.7&81.6&62.1&82.9&82.1 \\ \hline
			ours (IoU) &84.3&\textbf{82.9}&\textbf{83.4}&81.7&\textbf{80.0}&75.4&71.8&91.9&81.0  &80.9&92.5&59.2&\textbf{93.5}&\textbf{86.3}&\textbf{74.3}&\textbf{90.3}&\textbf{86.4} \\ \hline

		\end{tabular}
	\end{adjustbox}

	\end{center}
\end{table*}

The COSEG~\cite{wang2012active} dataset is also a commonly used benchmark for shape segmentation. Compared to ShapeNet, COSEG is much smaller. It has 8 scanned categories and 3 synthetic categories. Each of the 8 categories has around 20 models, which are too few for deep learning. The 3 synthetic categories each have 900 models, so we test our algorithm with the synthetic categories. We compare our result with \cite{xie20143d} and \cite{wang20183d} in Tab.~\ref{tab:coseg}. Our approach outperforms both of them.

\begin{table}[t]
	\begin{center}
	
			\caption{Classification accuracy on ModelNet10, ModelNet40 and ShapeNet. Distinguished by input representations, SPH~\cite{kazhdan2003rotation}, SyncSpecCNN~\cite{yi2017syncspeccnn}, ACNN~\cite{boscaini2016learning} and
			FoldingNet~\cite{boscaini2016learning} use meshes; PointNet~\cite{qi2017pointnet}, PointNet++\cite{qi2017pointnet} and SO-Net~\cite{li2018so} use point clouds; Voxelnet~\cite{maturana2015voxnet} and 3DShapeNets~\cite{wu20153d} take voxels as input. `-' in the table indicates the performance is not reported.
			For the classification accuracy on ShapeNet, our LaplacianNet achieves state-of-the-art performance. LaplacianNet outperforms all those single model classification methods.}
		\label{tab:class}
		\begin{tabular}{c|c|ccc}
			\hline
			Method&input&MN10&MN40&ShapeNet \\ \hline
			PointNet&point&-&89.2\%&-\\ 
			PointNet++&point&-&91.9\%&-\\
			SO-Net&point&95.7\%&93.4\%&-\\ \hline
			3DShapeNets&volume&83.5\%&77.0\%&-\\ 
			Voxnet&volume&91.0\%&84.5\%&-\\ \hline
			ACNN&mesh&-&-& 93.99\%\\
			SyncSpecCNN&mesh&-&-& 99.71\%\\
			SPH&mesh&-&68.2\%&-\\ \hline
			Ours&mesh&\textbf{97.4\%}&\textbf{94.21\%}&\textbf{99.88\%}\\ \hline
		\end{tabular}
	\end{center}
\end{table}
\begin{figure}[t]
	\begin{center}
		\includegraphics[width=\linewidth]{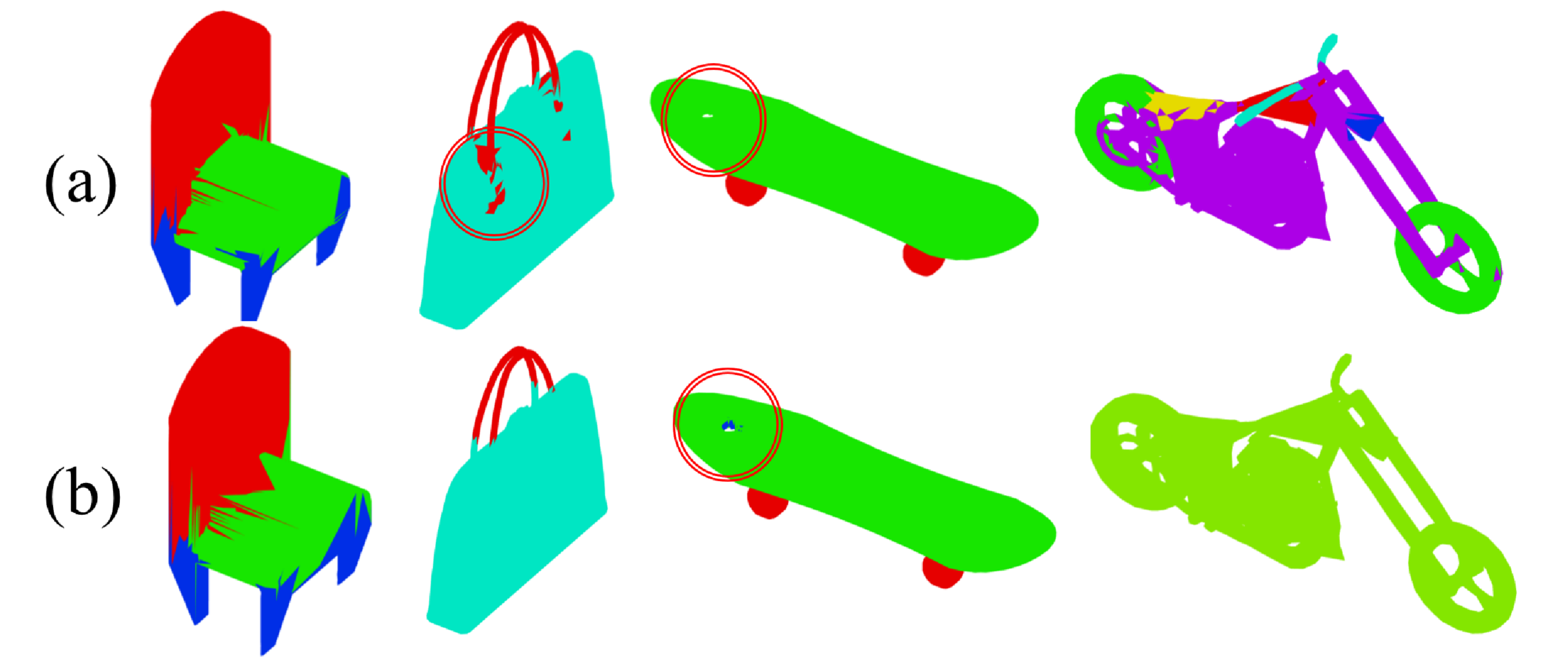}
	\end{center}
	\caption{Failure cases in ShapeNet segmentation. We show poor segmentation results in four categories in ShapeNet. (a) is the ground truth and (b) is our prediction. Those imperfect results are not necessarily caused by our segmentation method but 
	also imperfect labeling in the dataset and error accumulation during preprocessing.  Following examples show main challenges. The ground truth segmentation mask of the chair is not smooth, which shows the limitation of the annotation. Some regions of segmentation annotation on the bag are incorrect. This kind of errors can be caused during transferring the mesh into a watertight manifold. Our method also suffers from lacking training examples and fails to produce correct results on the motorbike. The skateboard has a hole in the mesh which leads to poor segmentation around it.}
	\label{fig:failure}
\end{figure}
\subsection{Mesh Classification}
In this section, we evaluate the accuracy of object categorization. The input features to this task are the same as the segmentation task. The application-specific network following Mesh Pooling Blocks is shown in Figure~\ref{fig:appnet}(bottom row). It contains two perceptron layers, a maxpooling layer and two fully connected layers. In this case the output is a softmax score for each category. We minimize the cross-entropy loss between one-hot vector of ground truth and network output. 

For this classification task, we use ModelNet10, ModelNet40~\cite{wu20153d} and ShapeNet. In each experiment, the network is trained for 200 epochs. Our method outperforms all state-of-the-art single model shape classification approaches in all the datasets.

\subsection{Failure Cases}
We would like to restate that our method does not work directly on the original ShapeNet models for two reasons: 1) the annotations are labeled on point clouds uniformly sampled from shapes, instead of vertices of the mesh; 2) meshes in ShapeNet are not manifold meshes, preventing us from performing high-quality spectral clustering. Therefore we convert shapes into watertight manifold surfaces using \cite{huang2018robust}, simplify the mesh to a reasonable level, and transfer part labels to vertices from the point cloud through nearest neighbor matching. Error accumulates during this process. In Figure~\ref{fig:failure} we show some failure cases when performing part segmentation in ShapeNet. There are several typical problems for models that are poorly segmented. In the chair example, we can observe sharp edges in ground truth labels, and such artifacts could be caused by the noise in the original point cloud. This kind of noise makes learning reasonable segmentation more challenging. As in the bag case, we can observe in the red circle that the ground truth is incorrect. This may be caused by changes of the shape while using \cite{huang2018robust} to make the model watertight, so that the nearest neighbor algorithm gets the wrong correspondence when transferring labels. We can see that our algorithm actually gets a more correct answer than the ``ground truth''. Failure in the motorbike case is caused by lack of training data. In the annotations, the motorbike class has the fewest data samples. Worse still, some of the training examples fail to be converted when performing \cite{huang2018robust}, because motorbikes have complex topological structure, making transfer challenging. In practice, it can be a big problem that errors in the transfer and simplification will decrease the amount of training data. Last but not least, the skateboard shows that there might be some holes, or other artifacts, in the surface, that could affect graph structure and mislead our algorithm.


\section{Conclusion}

In this paper, we present a deep learning approach to predicting functions defined on shapes. The key idea is to perform multiscale pooling based on Laplacian spectral clustering, and use a Correlation Net following pooling  to fuse global information. Compared to the pooling operation in the previous literature, our network does not require a uniform number of vertices in each model. 
Our work outperforms state-of-the-art methods in most categories for shape classification and segmentation. 

Several future extensions can be explored. First our LaplacianNet may be applied to other tasks. For example, 3D reconstruction  is fundamental but challenging. A capable and general method to generate meshes is of great demand. Our network has the potential to achieve this task, because it can neatly encode connectivity in vertices, and intrinsically understands the topology through spectral clustering and spatial pooling. Furthermore, as a general structure for mesh processing, our network may also be applied to shape deformation, completion and correspondence. 

Another area of future work is to design a kernel with trainable parameters, like convolutional kernels in images. For instance, a Gaussian kernel can be calculated inside a pooling cluster. Meanwhile, the transpose convolution plays an important role in traditional CNN frameworks. The combination of forward and transpose convolutions lets the network freely down-sample and up-sample data. 

Finally it would be interesting to see how our network can work on general graphs. In this paper, we mainly deal with manifold meshes, using geodesic distances to construct Laplacian. However in other problem settings, we can use any distance metric that can best describe the problem. For example, we might experiment with our method on social networks with arbitrary size and connectivity.

\ifCLASSOPTIONcompsoc
  \section*{Acknowledgments}
\else
  \section*{Acknowledgment}
\fi

This work was supported by National Natural Science Foundation of China~(No. 61872440 and No. 61828204), Beijing Natural Science Foundation~(No. L182016), Young Elite Scientists Sponsorship Program by CAST~(No. 2017QNRC001), Youth Innovation Promotion Association CAS, Huawei HIRP Open Fund~(No. HO2018085141), CCF-Tencent Open Fund, SenseTime Research Fund and Open Project Program of the National Laboratory of Pattern Recognition (NLPR).

\ifCLASSOPTIONcaptionsoff
  \newpage
\fi



%

\bibliographystyle{IEEEtran}
\bibliography{IEEEabrv,./egbib}

\begin{thebibliography}{10}
\providecommand{\url}[1]{#1}
\csname url@samestyle\endcsname
\providecommand{\newblock}{\relax}
\providecommand{\bibinfo}[2]{#2}
\providecommand{\BIBentrySTDinterwordspacing}{\spaceskip=0pt\relax}
\providecommand{\BIBentryALTinterwordstretchfactor}{4}
\providecommand{\BIBentryALTinterwordspacing}{\spaceskip=\fontdimen2\font plus
\BIBentryALTinterwordstretchfactor\fontdimen3\font minus
  \fontdimen4\font\relax}
\providecommand{\BIBforeignlanguage}[2]{{%
\expandafter\ifx\csname l@#1\endcsname\relax
\typeout{** WARNING: IEEEtran.bst: No hyphenation pattern has been}%
\typeout{** loaded for the language `#1'. Using the pattern for}%
\typeout{** the default language instead.}%
\else
\language=\csname l@#1\endcsname
\fi
#2}}
\providecommand{\BIBdecl}{\relax}
\BIBdecl

\bibitem{karpathy2014large}
A.~Karpathy, G.~Toderici, S.~Shetty, T.~Leung, R.~Sukthankar, and L.~Fei-Fei,
  ``Large-scale video classification with convolutional neural networks,'' in
  \emph{Proceedings of the IEEE conference on Computer Vision and Pattern
  Recognition}, 2014, pp. 1725--1732.

\bibitem{chen2017reference}
M.~Chen, G.~Ding, S.~Zhao, H.~Chen, Q.~Liu, and J.~Han, ``Reference based
  {LSTM} for image captioning.'' in \emph{AAAI}, 2017, pp. 3981--3987.

\bibitem{long2015fully}
J.~Long, E.~Shelhamer, and T.~Darrell, ``Fully convolutional networks for
  semantic segmentation,'' in \emph{Proceedings of the IEEE conference on
  computer vision and pattern recognition}, 2015, pp. 3431--3440.

\bibitem{kwak2017weakly}
S.~Kwak, S.~Hong, B.~Han \emph{et~al.}, ``Weakly supervised semantic
  segmentation using superpixel pooling network.'' in \emph{AAAI}, 2017, pp.
  4111--4117.

\bibitem{wu20153d}
Z.~Wu, S.~Song, A.~Khosla, F.~Yu, L.~Zhang, X.~Tang, and J.~Xiao, ``3d
  shapenets: A deep representation for volumetric shapes,'' in
  \emph{Proceedings of the IEEE conference on computer vision and pattern
  recognition}, 2015, pp. 1912--1920.

\bibitem{bronstein2017geometric}
M.~M. Bronstein, J.~Bruna, Y.~LeCun, A.~Szlam, and P.~Vandergheynst,
  ``Geometric deep learning: going beyond {E}uclidean data,'' \emph{IEEE Signal
  Processing Magazine}, vol.~34, no.~4, pp. 18--42, 2017.

\bibitem{masci2015geodesic}
J.~Masci, D.~Boscaini, M.~Bronstein, and P.~Vandergheynst, ``Geodesic
  convolutional neural networks on {R}iemannian manifolds,'' in
  \emph{Proceedings of the IEEE international conference on computer vision
  workshops}, 2015, pp. 37--45.

\bibitem{boscaini2016learning}
D.~Boscaini, J.~Masci, E.~Rodol{\`a}, and M.~Bronstein, ``Learning shape
  correspondence with anisotropic convolutional neural networks,'' in
  \emph{Advances in Neural Information Processing Systems}, 2016, pp.
  3189--3197.

\bibitem{Crane:2017:HMD}
\BIBentryALTinterwordspacing
K.~Crane, C.~Weischedel, and M.~Wardetzky, ``The heat method for distance
  computation,'' \emph{Commun. ACM}, vol.~60, no.~11, pp. 90--99, Oct. 2017.
  [Online]. Available: \url{http://doi.acm.org/10.1145/3131280}
\BIBentrySTDinterwordspacing

\bibitem{2013spectral}
J.~Bruna, W.~Zaremba, A.~Szlam, and Y.~LeCun, ``Spectral networks and locally
  connected networks on graphs,'' \emph{arXiv preprint arXiv:1312.6203}, 2013.

\bibitem{li2018adaptive}
R.~Li, S.~Wang, F.~Zhu, and J.~Huang, ``Adaptive graph convolutional neural
  networks,'' \emph{arXiv preprint arXiv:1801.03226}, 2018.

\bibitem{yi2017syncspeccnn}
L.~Yi, H.~Su, X.~Guo, and L.~J. Guibas, ``{SyncSpecCNN}: Synchronized spectral
  {CNN} for 3d shape segmentation.'' in \emph{CVPR}, 2017, pp. 6584--6592.

\bibitem{qi2016volumetric}
C.~R. Qi, H.~Su, M.~Nie{\ss}ner, A.~Dai, M.~Yan, and L.~J. Guibas, ``Volumetric
  and multi-view {CNNs} for object classification on 3d data,'' in \emph{CVPR},
  2016, pp. 5648--5656.

\bibitem{qi2017pointnet}
C.~R. Qi, H.~Su, K.~Mo, and L.~J. Guibas, ``{PointNet}: Deep learning on point
  sets for 3d classification and segmentation,'' \emph{Proc. Computer Vision
  and Pattern Recognition (CVPR), IEEE}, vol.~1, no.~2, p.~4, 2017.

\bibitem{qi2017pointnet++}
C.~R. Qi, L.~Yi, H.~Su, and L.~J. Guibas, ``{PointNet}++: Deep hierarchical
  feature learning on point sets in a metric space,'' in \emph{Advances in
  Neural Information Processing Systems}, 2017, pp. 5099--5108.

\bibitem{henaff2015deep}
M.~Henaff, J.~Bruna, and Y.~LeCun, ``Deep convolutional networks on
  graph-structured data,'' \emph{arXiv preprint arXiv:1506.05163}, 2015.

\bibitem{kalogerakis2010learning}
E.~Kalogerakis, A.~Hertzmann, and K.~Singh, ``Learning 3d mesh segmentation and
  labeling,'' \emph{ACM Transactions on Graphics (TOG)}, vol.~29, no.~4, p.
  102, 2010.

\bibitem{sun2009concise}
J.~Sun, M.~Ovsjanikov, and L.~Guibas, ``A concise and provably informative
  multi-scale signature based on heat diffusion,'' in \emph{Computer graphics
  forum}, vol.~28, no.~5.\hskip 1em plus 0.5em minus 0.4em\relax Wiley Online
  Library, 2009, pp. 1383--1392.

\bibitem{aubry2011wave}
M.~Aubry, U.~Schlickewei, and D.~Cremers, ``The wave kernel signature: A
  quantum mechanical approach to shape analysis,'' in \emph{Computer Vision
  Workshops (ICCV Workshops), 2011 IEEE International Conference on}.\hskip 1em
  plus 0.5em minus 0.4em\relax IEEE, 2011, pp. 1626--1633.

\bibitem{von2007tutorial}
U.~Von~Luxburg, ``A tutorial on spectral clustering,'' \emph{Statistics and
  computing}, vol.~17, no.~4, pp. 395--416, 2007.

\bibitem{wu2013unsupervised}
Z.~Wu, Y.~Wang, R.~Shou, B.~Chen, and X.~Liu, ``Unsupervised co-segmentation of
  3d shapes via affinity aggregation spectral clustering,'' \emph{Computers \&
  Graphics}, vol.~37, no.~6, pp. 628--637, 2013.

\bibitem{shu2016unsupervised}
Z.~Shu, C.~Qi, S.~Xin, C.~Hu, L.~Wang, Y.~Zhang, and L.~Liu, ``Unsupervised 3d
  shape segmentation and co-segmentation via deep learning,'' \emph{Computer
  Aided Geometric Design}, vol.~43, pp. 39--52, 2016.

\bibitem{huang2011joint}
Q.~Huang, V.~Koltun, and L.~Guibas, ``Joint shape segmentation with linear
  programming,'' \emph{ACM Transactions on Graphics (TOG)}, vol.~30, no.~6, p.
  125, 2011.

\bibitem{sidi2011unsupervised}
O.~Sidi, O.~van Kaick, Y.~Kleiman, H.~Zhang, and D.~Cohen-Or,
  \emph{Unsupervised co-segmentation of a set of shapes via descriptor-space
  spectral clustering}.\hskip 1em plus 0.5em minus 0.4em\relax ACM, 2011,
  vol.~30, no.~6.

\bibitem{chen2009benchmark}
X.~Chen, A.~Golovinskiy, and T.~Funkhouser, ``A benchmark for 3d mesh
  segmentation,'' \emph{ACM Transactions on Graphics (TOG)}, vol.~28, no.~3,
  p.~73, 2009.

\bibitem{yi2016scalable}
L.~Yi, V.~G. Kim, D.~Ceylan, I.~Shen, M.~Yan, H.~Su, C.~Lu, Q.~Huang,
  A.~Sheffer, L.~Guibas \emph{et~al.}, ``A scalable active framework for region
  annotation in 3d shape collections,'' \emph{ACM Transactions on Graphics
  (TOG)}, vol.~35, no.~6, p. 210, 2016.

\bibitem{wang2012active}
Y.~Wang, S.~Asafi, O.~Van~Kaick, H.~Zhang, D.~Cohen-Or, and B.~Chen, ``Active
  co-analysis of a set of shapes,'' \emph{ACM Transactions on Graphics (TOG)},
  vol.~31, no.~6, p. 165, 2012.

\bibitem{shilane2004princeton}
P.~Shilane, P.~Min, M.~Kazhdan, and T.~Funkhouser, ``The {P}rinceton shape
  benchmark,'' in \emph{Shape modeling applications, 2004. Proceedings}.\hskip
  1em plus 0.5em minus 0.4em\relax IEEE, 2004, pp. 167--178.

\bibitem{george20183d}
D.~George, X.~Xie, and G.~K. Tam, ``3d mesh segmentation via multi-branch 1d
  convolutional neural networks,'' \emph{Graphical Models}, vol.~96, pp. 1--10,
  2018.

\bibitem{wang20183d}
P.~Wang, Y.~Gan, P.~Shui, F.~Yu, Y.~Zhang, S.~Chen, and Z.~Sun, ``3d shape
  segmentation via shape fully convolutional networks,'' \emph{Computers \&
  Graphics}, vol.~70, pp. 128--139, 2018.

\bibitem{kalogerakis20173d}
E.~Kalogerakis, M.~Averkiou, S.~Maji, and S.~Chaudhuri, ``3d shape segmentation
  with projective convolutional networks,'' in \emph{Proc. CVPR}, vol.~1,
  no.~2, 2017, p.~8.

\bibitem{guo20153d}
K.~Guo, D.~Zou, and X.~Chen, ``3d mesh labeling via deep convolutional neural
  networks,'' \emph{ACM Transactions on Graphics (TOG)}, vol.~35, no.~1, p.~3,
  2015.

\bibitem{Meyer2003}
M.~Meyer, M.~Desbrun, P.~Schr\"{o}der, and A.~H. Barr, ``Discrete
  differential-geometry operators for triangulated 2-manifolds,'' in
  \emph{Visualization and Mathematics III}, 2003, pp. 35--57.

\bibitem{rumelhart1985learning}
D.~E. Rumelhart, G.~E. Hinton, and R.~J. Williams, ``Learning internal
  representations by error propagation,'' California Univ San Diego La Jolla
  Inst for Cognitive Science, Tech. Rep., 1985.

\bibitem{kingma2014adam}
D.~P. Kingma and J.~Ba, ``Adam: A method for stochastic optimization,''
  \emph{arXiv preprint arXiv:1412.6980}, 2014.

\bibitem{Garland1997}
\BIBentryALTinterwordspacing
M.~Garland and P.~S. Heckbert, ``Surface simplification using quadric error
  metrics,'' in \emph{Proceedings of the 24th Annual Conference on Computer
  Graphics and Interactive Techniques}, ser. SIGGRAPH '97.\hskip 1em plus 0.5em
  minus 0.4em\relax New York, NY, USA: ACM Press/Addison-Wesley Publishing Co.,
  1997, pp. 209--216. [Online]. Available:
  \url{https://doi.org/10.1145/258734.258849}
\BIBentrySTDinterwordspacing

\bibitem{chang2015shapenet}
A.~X. Chang, T.~Funkhouser, L.~Guibas, P.~Hanrahan, Q.~Huang, Z.~Li,
  S.~Savarese, M.~Savva, S.~Song, H.~Su \emph{et~al.}, ``{ShapeNet}: An
  information-rich 3d model repository,'' \emph{arXiv preprint
  arXiv:1512.03012}, 2015.

\bibitem{huang2018robust}
J.~Huang, H.~Su, and L.~Guibas, ``Robust watertight manifold surface generation
  method for {ShapeNet} models,'' \emph{arXiv preprint arXiv:1802.01698}, 2018.

\bibitem{verma2018feastnet}
N.~Verma, E.~Boyer, and J.~Verbeek, ``{FeaStNet}: Feature-steered graph
  convolutions for 3d shape analysis,'' in \emph{CVPR 2018-IEEE Conference on
  Computer Vision \& Pattern Recognition}, 2018.

\bibitem{xie20143d}
Z.~Xie, K.~Xu, L.~Liu, and Y.~Xiong, ``3d shape segmentation and labeling via
  extreme learning machine,'' in \emph{Computer graphics forum}, vol.~33,
  no.~5.\hskip 1em plus 0.5em minus 0.4em\relax Wiley Online Library, 2014, pp.
  85--95.

\bibitem{kazhdan2003rotation}
M.~Kazhdan, T.~Funkhouser, and S.~Rusinkiewicz, ``Rotation invariant spherical
  harmonic representation of 3d shape descriptors,'' in \emph{Symposium on
  geometry processing}, vol.~6, 2003, pp. 156--164.

\bibitem{li2018so}
J.~Li, B.~M. Chen, and G.~H. Lee, ``{SO-Net}: Self-organizing network for point
  cloud analysis,'' in \emph{Proceedings of the IEEE Conference on Computer
  Vision and Pattern Recognition}, 2018, pp. 9397--9406.

\bibitem{maturana2015voxnet}
D.~Maturana and S.~Scherer, ``Voxnet: A 3d convolutional neural network for
  real-time object recognition,'' in \emph{Intelligent Robots and Systems
  (IROS), 2015 IEEE/RSJ International Conference on}.\hskip 1em plus 0.5em
  minus 0.4em\relax IEEE, 2015, pp. 922--928.

\end{thebibliography}


%









\end{document}